\documentclass[aps,pre,superscriptaddress,twocolumn,showpacs,floatfix]{revtex4}
\usepackage{graphics,dcolumn,amsmath,amssymb}
\usepackage{epsfig}
\begin{document}
                                                        
\title{Correlations in Networks associated to Preferential Growth}
\author{Andreas  \surname{Gr\"{o}nlund}}
\email{gronlund@tp.umu.se}
\affiliation{Department of Theoretical Physics, Ume{\aa} University,
901 87 Ume{\aa}, Sweden}

\author{Kim \surname{Sneppen}}
\affiliation{Niels Bohr Institute, Blegdamsvej 17, DK-2100 Copenhagen, Denmark}

\author{Petter \surname{Minnhagen}}
\affiliation{Department of Theoretical Physics, Ume{\aa} University,
901 87 Ume{\aa}, Sweden}
\affiliation{NORDITA, Blegdamsvej 17, DK-2100 Copenhagen, Denmark}

\begin{abstract}
Combinations of random and preferential growth for both on-growing and stationary networks are studied and a hierarchical topology is observed. 
Thus for real world scale-free networks which do not exhibit hierarchical features preferential growth is probably not the main ingredient in 
the growth process. An example of such real world networks includes the protein-protein interaction network in yeast, which exhibits pronounced 
anti-hierarchical features.
\end{abstract}

\pacs{89.75.-k, 89.70.+c, 05.10.-a, 05.65.+b}

\maketitle

One feature that many complex networks show is scale-free degree distribution of vertices, 
that is the probability of finding a vertex of degree $k$ follows 
$P(k)\propto k^{-\gamma}$. A popular explanation for the scale-free degree 
distribution of vertices is preferential attachment \cite{dsprice2,ba:model} in which new 
vertices tend to connect themselves to already highly connected vertices. 
In addition to the degree distribution there are additional topological measures
that can be used to characterize networks, for example degree-degree correlations, 
that is ``who is connected to who?''. An understanding of by what
process networks emerge should then include an understanding of the corresponding topological
 measures both for real networks and for networks models \cite{redner:sitges,mejn:oricorr,maslov:pro,mejn:clupref}. 
In particular it has been observed that protein-protein networks have quite different degree-degree 
correlations than the Internet \cite{maslov:inet}, although both molecular networks and the Internet show scale-free features.
In the  present paper we investigate versions of preferential attachment both for on-growing and stationary networks, 
and study the degree distribution and the degree-degree-correlations. Our conclusion is that preferential attachment 
is robust with respect to a hierarchical type of degree-degree correlations. As a consequence, real networks which do not 
have this type of degree-degree correlations are unlikely to have evolved by a version of preferential attachment.
\\

A network, or more formally a graph, $G(V,E)$ consists of a set of vertices $V$ and a 
set of edges $E$ which connect pairs of vertices in the network. It can both be ordered
and unordered pairs depending if the network is directed or not. 
We only consider undirected networks here. 
When generating such a network we consider four elementary processes: 
addition or removal of respectively vertices or edges. 
Here we use preferential attachment when adding new vertices or edges to the graph, 
either preferential attachment in itself or combined with random attachment.
We will furthermore consider both a growing network, and a non growing 
network evolving by addition and removal of vertices 
and edges at steady state conditions. 

\section{Growing Networks}

First let us consider a network grown to some number of vertices $N$, that we fix 
from the beginning (typically we use $N=10^3$).
The network grows to this size by a process where we at each step 
do the following:

\begin{itemize}
\item With probability $p$ a new vertex is added and connected with an edge to a preferentially selected vertex.
\item With probability $1-p$ a new edge is added between two vertices which are
  \begin{itemize}
  \item with probability $q$ both chosen preferentially.
  \item with probability $1-q$ one vertex is chosen preferentially and the other vertex is randomly chosen.
  \end{itemize}
\end{itemize}

Double edges or loops are not allowed and therefore each time we add an edge 
to the network, a check is performed.
If the connection is not valid, one attempts to put the edge somewhere else.
This will always be possible, except for some non important cases where the network
is very small. To have good statistics a number of networks are grown to the desired size $N$ by
the rules above. Also for every network that is produced one makes 
a sample of randomized networks with exactly the same degree 
distribution as the grown network, as described in \cite{maslov:pro}. 
We look at the like-hood of having a connection between vertices of
edge-degree $K_1$ to vertices of edge-degree $K_2$ in the real network
and compare it with the probability of finding the same connection 
obtained in the random sample of network \cite{maslov:pro}:
\begin{equation}
R(K_1,K_2) \; = \; \frac{P(K_1,K_2)}{P_{random} (K_1,K_2)}
\end{equation}
The reason for comparing with a set of rewired networks is because of the
inherently complicated nature of a network. 
So far the analytical approaches only applies to networks where multiple edges between two vertices and loops are allowed.
With a specific degree distribution and not allowing for loops or
multiple edges between vertices there is a limited freedom when
attaching edges.
This restriction will give a preference to small vertices connecting
with large vertices in a scale-free network. 
In order to measure how the correlations in the created network differs from the one 
expected from a network with the same degree sequence we divide the number of specific 
connections in the studied network with the number of connections in the randomized networks.
In principle one could also have obtained information about this ``two-point''
correlation from the measure of assortative mixing \cite{mejn:assmix}, 
compared with the set of randomized networks, 
but the full correlation profile contains more specific 
details about the topology.

\begin{figure}[htb]
\begin{center}
\includegraphics[width=0.95\columnwidth]{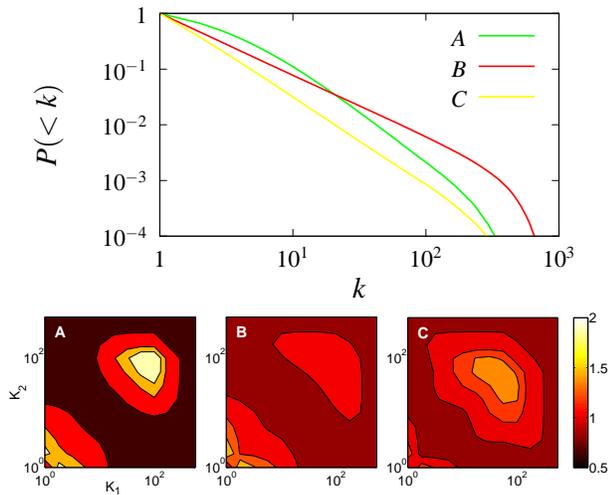}
\caption{Growing networks. Cumulative plot of the degree distributions $P(<k)$ for the networks on top and below the correlation profiles $R(K_1,K_2)$ for the respective networks. The number of vertices are $1000$ for the correlation profiles and $10000$ for the degree distributions. The different networks above are generated with the parameters: ({\bf A}) $p=0.4$ and $q=0$. ({\bf B}) $p=0.4$ and $q=1$. ({\bf C}) $p=0.8$ and $q=1$.}
\label{Fig1}
\end{center}
\end{figure}

In the figure \ref{Fig1} we show the degree distribution of three differently grown networks labeled
A, B and C. That is we consider growth with different rates of edge additions to vertex additions
as quantified by $p$. Further, the given $p=0.4$ corresponds to adding 4 
vertices each with one edge attached preferentially, to every elementary addition of
6 edges. The $q=1$ and $q=0$ corresponds to preferential attachment of 
these 6 edges in both ends, respectively to attachment of one of their ends 
to a randomly selected vertex.
In all cases one obtains scale free networks \cite{doro:cont,cald:gener}:
\begin{equation}
P(k) \propto \frac{1}{k^{\gamma}}
\end{equation}
with exponent $\gamma$ that decreases with both $p$ and $q$. For the three cases in figure \ref{Fig1} we have that (A) $p=0.4$ with $q=0$ gives $\gamma=2.86$, (B) $p=0.4$ with $q=1$ gives $\gamma=2.14$ and (C) $p=0.8$ with $q=1$ gives $\gamma=2.6$.

Figure \ref{Fig1} (A-C) examines the correlation profile.
The overall pattern is that in all cases highly connected vertices tend to
connect to highly connected vertices, a feature which in \cite{trusina:hier} was associated with
 hierarchical topologies of networks.
Also an overall pattern, is that the more edges there are, the more $R(K_1,K_2)$ approaches unity 
and the hierarchical topologies thus tend to be suppressed by the
overall noise. Examining the different types of growth, we furthermore see that the most hierarchical 
networks are obtained when edges are added randomly in one end and preferentially 
in the other end. In a somewhat similar vein assortativity was studied
in \cite{cald:assmo}.

\begin{figure}[htb]
\begin{center}
\includegraphics[width=0.95\columnwidth]{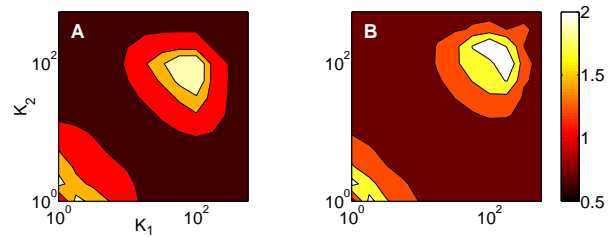}
\caption{Comparison of correlation profile for a network process ({\bf
    A}) $p=0.4$ and $q=0$, and the same process with the only
  difference that double edges and loops are allowed, ({\bf B}).}
\label{Fig2}
\end{center}
\end{figure}

Since analytical calculations usually are limited to only apply to networks where loops and multiple 
edges between vertices are allowed, we also investigate what effect this has to the correlation profile 
by performing the same growth process as before, but with the difference that multiple edges and 
loops are accepted.
Multiple edges and loops are accepted both in the process of growing the networks and the creation
of the randomized networks. In figure \ref{Fig2} the correlation profile is visible.

\begin{figure}[htb]
\begin{center}
\includegraphics[width=0.95\columnwidth]{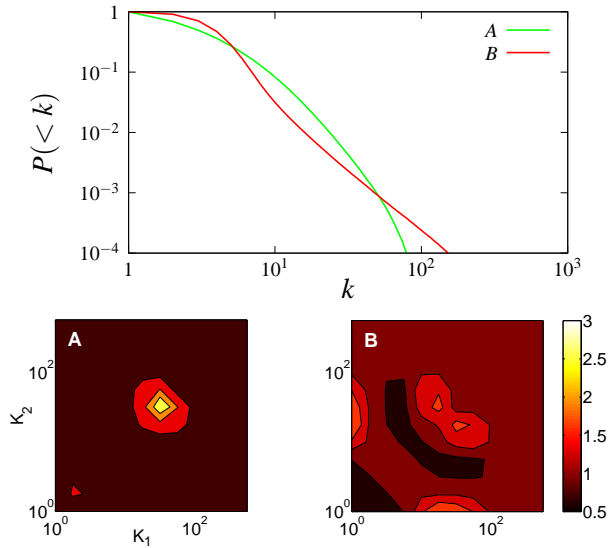}
\caption{Cumulative plot of the degree distributions $P(<k)$ for two
  network processes generating a network of $N=1000$ vertices. 
  ({\bf A}) is a process of preferentially vertex
  and edge insertions, with $p=0.5$. The edges are inserted
  randomly in both ends. ({\bf B}) a process where the excess
  edges are inserted randomly after all the vertices are 
  inserted preferentially to the network. 
  The number of edges are the same for the two networks, $M=2000$}
\label{Fig3}
\end{center}
\end{figure}

In the figure \ref{Fig2} we see that, even if double edges and loops are allowed,
 indeed the highly connected vertices are connected more frequently to each other 
compared with a maximal randomization.
However, the peak is shifted towards higher degrees because many edges are allowed between two vertices.
If still more edges to vertices are inserted, 
the differences will be even larger because of the number of double 
edges and loops that will be created.
The preferential attachment is however not the full story of the
 correlations, there are more to it.
In the process of preferential attachment, 
the oldest vertices tend to become the vertices of highest degree.
Furthermore, the insertion of an edge in the network connects two vertices 
created before the time of the edge insertion.
This implicates that when the network is created and all vertex and edge
 insertions are made, more edges are put between older vertices than the
younger vertices simply because the network is smaller in the early stages
 of the growth process; thus older vertices have a higher probability 
to be connected by an edge than the younger vertices created in the later stages.
This explains why even if the edges are inserted randomly in both
 ends one gets a highly hierarchical structure, figure \ref{Fig3}A, 
compared to what is expected from the resulting degree distribution.
The degree distribution no longer follows a power law but is still
 fairly broad. 
Comparing to the the process where the excess edges are inserted
 after the insertion of all the vertices to the network, figure
 \ref{Fig3}B, we observe that the hierarchical structure is no longer
 as apparent as in figure \ref{Fig3}A. Furthermore, as a consequence of
 inserting fewer edges to the older (more connected) vertices the degree
 distribution is not as broad as in \ref{Fig3}A.

\section{Stationary Networks}
Many real world networks are not constantly growing, but may anyway be governed
by a growth process, that then should be supplemented by means of elimination
of parts of the network.
In the case of preferential growth the oldest vertices also become the most 
central ones which is shown in the degree-degree correlation profile \ref{Fig1}. 
It is therefore of interest to examine what happens if the oldest vertices may 
be randomly eliminated. This is investigated in the following
steady state model for growth and elimination in networks.
\\

Given we want a network consisting of $N$ vertices, we grow the network as before,
but in addition add a removal step at any time the number of vertices exceeds $N$.
The total algorithm then reads:

\begin{itemize}
\item With probability $p$ a new vertex is added and connected with an edge to a preferentially selected vertex.
\item With probability $1-p$ a new edge is added between two vertices which are
  \begin{itemize}
  \item with probability $q$ both chosen preferentially.
  \item with probability $1-q$ one vertex is chosen preferentially and the other vertex randomly chosen.
  \end{itemize}
\item If \#vertices $>$ N, remove a random vertex $n$ and all vertices that after the removal of $n$ becomes isolated.
\end{itemize}

\begin{figure}[htb]
\begin{center}
\includegraphics[width=0.95\columnwidth]{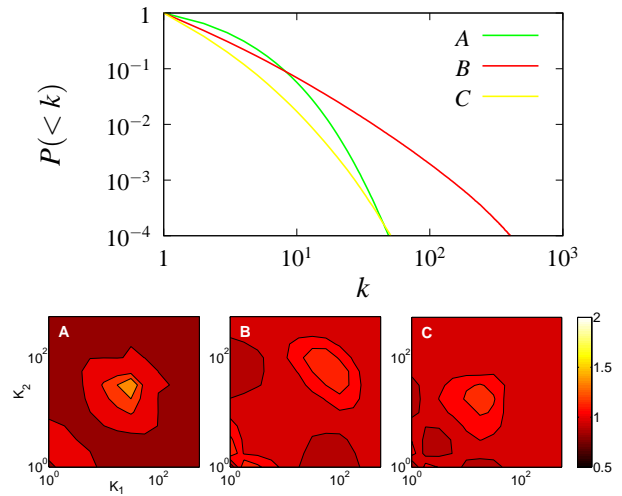}
\caption{Stationary networks. Cumulative plot of the degree distributions $P(<K)$ for the networks on top and below the correlation profiles $R(K_1,K_2)$ for the respective networks. The number of vertices are $1000$ for the correlation profiles and $10000$ for the degree distributions. The different networks above are generated with the parameters: ({\bf A}) $p=0.4$ and $q=0$. ({\bf B}) $p=0.4$ and $q=1$. ({\bf C}) $p=0.8$ and $q=1$.}
\label{Fig4}
\end{center}
\end{figure}

At given time-steps (typically at the order of the size of the network), 
randomizations of the network are made in order to calculate the 
degree-degree correlation profiles.
Figure \ref{Fig4} demonstrates that now, with both growth and elimination, 
the scale invariance is broken. 
This is a striking difference to the original Simon \cite{ha2} model of ``rich get richer''.
In his model money was assigned to people stochastically 
with a probability given by their present wealth leading to a power law distribution of wealth.

In his case one also obtains a power law distribution if one randomly
 eliminates agents independently of their wealth, see also 
\cite{bornholdt:WWW}. The reason for the different behavior in 
the network case is due to the fact that when one eliminates vertices 
with few connections, then with high probability one also reduces the 
number of connections for the vertices with high degrees.

Considering the correlation profiles for the steady state networks one
first of all notices that hierarchical features remain. Further,
when comparing to the steadily growing networks the hierarchical 
features are suppressed.
Also notice that for the high $p$ or low $q$ the degree distribution 
became close to exponential, a feature that in itself will diminish the
importance of the edge degree as an informative characteristic of the vertex structure.
However, the relative strength of observed correlations for steady state networks are qualitatively similar to what was obtained for the growing networks.
\\

In summary we have shown that preferential attachment and continuous
edge insertions leads to a rather robust characteristic type of
hierarchical degree-degree correlations. 
Thus for real world scale-free networks that does not exhibit hierarchical features, 
preferential growth is probably not the main ingredient in forming their topology.
An example of such real world networks includes the protein-protein interaction networks in yeast, which exhibits pronounced 
anti-hierarchical topology \cite{maslov:pro,trusina:hier}.
Thus the robustness of the hierarchical topology that preferential attachment gives rise 
to, points to some difficulty in the preferential attachment 
scenario put forward in \cite{eli:pro-pref} for protein-protein networks,
 not withstanding the fact that it was found that the older proteins were observed to 
be more connected than the younger ones. 

\bibliographystyle{apsrev}
\bibliography{mybib}

\end{document}